\documentclass[runningheads]{llncs}
\usepackage{graphicx}
\usepackage{subfigure} 
\usepackage{multirow}
\usepackage{array}%需要该宏包
\usepackage{booktabs}
\usepackage{threeparttable}
\usepackage{amsmath}
\usepackage{amsfonts}
\usepackage{amssymb}
\usepackage{bm}
\usepackage{multicol}
\usepackage[hidelinks]{hyperref}
\usepackage{float}
\usepackage[misc]{ifsym}

\begin{document}

\title{XCrossNet: Feature Structure-Oriented Learning \\ for Click-Through Rate Prediction}
\titlerunning{XCrossNet: Feature Structure-Oriented Learning for CTR prediction}
% If the paper title is too long for the running head, you can set
% an abbreviated paper title here
%
\author{Runlong Yu\inst{1} \and Yuyang Ye\inst{2} \and Qi Liu\inst{1(\text{\Letter})} \and Zihan Wang\inst{3} \and Chunfeng Yang\inst{4} \and Yucheng Hu\inst{4} \and Enhong Chen\inst{1}}

\authorrunning{Runlong Yu, et al.}
% First names are abbreviated in the running head.
% If there are more than two authors, 'et al.' is used.

\institute{Anhui Province Key Laboratory of Big Data Analysis and Application, \\ School of Computer Science and Technology, \\ University of Science and Technology of China, Hefei, China \\
	\email{yrunl@mail.ustc.edu.cn, \{qiliuql,cheneh\}@ustc.edu.cn}
	\and
Management Science and Information Systems, Rutgers Business School, \\ Rutgers University, Newark, USA
\email{yuyang.ye@rutgers.edu}\\ \and
MOE Key Laboratory of Computational Linguistics, \\
School of Electronics Engineering and Computer Science, \\ Peking University, Beijing, China 
\email{wzh@stu.pku.edu.cn} \and
Tencent Inc, Shenzhen, China 
\email{\{yannisyang,nikohu\}@tencent.com}}
\maketitle              % typeset the header of the contribution
\begin{abstract}
	Click-Through Rate (CTR) prediction is a core task in nowadays commercial recommender systems. Feature crossing, as the mainline of research on CTR prediction, has shown a promising way to enhance predictive performance.
	Even though various models are able to learn feature interactions without manual feature engineering, they rarely attempt to individually learn representations for different feature structures.
	In particular, they mainly focus on the modeling of cross sparse features but neglect to specifically represent cross dense features. 
	Motivated by this, we propose a novel Extreme Cross Network, abbreviated XCrossNet, which aims at learning dense and sparse feature interactions in an explicit manner. 
	XCrossNet as a feature structure-oriented model leads to a more expressive representation and a more precise CTR prediction, which is not only explicit and interpretable, but also time-efficient and easy to implement.
	Experimental studies on Criteo Kaggle dataset show significant improvement of XCrossNet over state-of-the-art models on both effectiveness and efficiency.
	
\end{abstract}

\section{Introduction}

Accurate targeting of commercial recommender systems is of great importance, in which Click-Through Rate~(CTR) prediction plays a key role.
CTR prediction aims to estimate the ratio of clicks to the impression of a recommended item for a user. 
Therefore, we consider users have negative preferences instead of implicit feedbacks on those un-clicked items \cite{yu2020collaborative,yu2018multiple}.
In common display advertising systems, advertisers expect lower costs to achieve a higher return on investment. The ad exchange platforms usually trade with advertisers and publishers according to the generalized second price of the maximum effective Cost Per Mille~(eCPM). If CTR is overestimated, advertisers could waste campaign budgets on the useless impression; On the other hand, if CTR is underestimated, advertisers would lose some valuable impressions and the campaigns may under deliver. 
With multi-billion dollar business on commercial recommendation today, CTR prediction has received growing interest from communities of both academia and industry \cite{cheng2016wide,liu2011personalized,wang2017deep}.

\addtocounter{footnote}{-4}

In web-scale commercial recommender systems, the inputs of users' characteristics are in two kinds of structures. The first kind of structure is described by numerical or dense parameters, e.g., ``\texttt{Age\_years=22, Height\_cm=165}''. Each of such characteristics is formalized as a value associated with a numerical field, while the values are named as dense features. The second kind of structure is described by categorical or sparse parameters, e.g.,``\texttt{Gender=Female, Relationship=In love}''. Each of such characteristics is formalized as a vector of one-hot encoding associated with a categorical field, while the vectors are named as sparse features. 
Research shows an important property of recommendation datasets for industrial use cases is the availability of both dense features and sparse features \cite{wu2020developing}. Thus, Criteo Kaggle dataset\footnote{https://labs.criteo.com/2013/12/download-terabyte-click-logs/} is usually regarded as representative of real production use cases. Moreover, the number of dense and sparse features for industrial use cases are often 100s to 1000 with a 50:50 split\footnote{The statistics for the dense and sparse features and proportion are based on the survey outcome conducted in December 2019 with the MLPerf Advisory Board.}.

Data scientists usually spend much time on interactions of raw features to generate better predictive models. 
Among these feature interactions, cross features, previously focused more on the product of sparse features,  show a promising way to enhance the performance of prediction~\cite{luo2019autocross}.
Owing to the fact that correct cross features are mostly task-specific and difficult to identify a priori, the crucial challenge is in automatically extracting sophisticated cross features hidden in high-dimensional data. 
Research on feature crossing as the mainline of CTR prediction has attracted widespread attention in recent years. Shallow models are simple, interpretable, and easy to scale, but limited in expressive ability. Alternatively, deep learning has shown powerful expressive capabilities, nevertheless, deep neural networks (DNNs) require many more parameters than tensor factorization to approximate high-order cross features.
Besides, almost all deep models leverage multilayer perceptron (MLP) to learn high-order feature interactions, however, whether plain DNNs indeed effectively represent right functions of cross features remains an open question~\cite{lian2018xdeepfm,wang2017deep}. 

 In addition, most methods neglect to represent cross dense features. There are three major patterns for handling dense features. First, dense features are discarded when crossing features, that is, dense features only participate in the linear part of the model \cite{DBLP:conf/icdm/Rendle10}. Second, dense features are directly concatenated with the embeddings of sparse features, which could cause an important feature dimensionality imbalance problem \cite{wang2017deep}. Third, dense features are converted into sparse features through bucketing, which could introduce hyper-parameters and loss information of dense features \cite{DBLP:conf/sigir/LiuXGTZHL20}.

Based on all these observations, we propose a novel \textbf{Extreme Cross Network (XCrossNet)}, to represent feature structure-oriented interactions. Modeling with XCrossNet consists of three stages. In the Feature Crossing stage, we separately propose a cross layer for crossing dense features and a product layer for crossing sparse features. In the Feature Concatenation stage, cross dense features and cross sparse features interact through a concatenate layer and a cross layer. Lastly, in the Feature Selection stage, we employ an MLP for capturing non-linear interactions and their relative importance.
Experimental results on Criteo Kaggle dataset demonstrate the superior performance of XCrossNet over the state-of-the-art baselines.

\section{Related Work}

Studies on CTR prediction can be categorized into five classes which will be respectively introduced below. 

\textbf{Generalized linear models.} Logistic Regression (LR) models such as FTRL are widely used in CTR prediction for their simplicity and efficiency~\cite{DBLP:conf/kdd/LeeODL12,DBLP:conf/kdd/McMahanHSYEGNPDGCLWHBK13}.
Ling Yan et al. argue that LR cannot capture nonlinear feature interactions and propose Coupled Group Lasso (CGL) to solve it \cite{DBLP:conf/icml/YanLXH14}. Human efforts are usually needed for LR models. 
Gradient Boosting Decision Tree (GBDT) is a method to automatically do feature engineering and search interactions \cite{friedman2001greedy}, then the transformed feature interactions can be fed into LR. In practice, tree-based models are more suitable for dense features but not for sparse features.

\textbf {Quadratic polynomial mappings and Factorization Machines.} Poly2 enumerates all pairwise feature interactions to avoid feature engineering which works well on dense features \cite{DBLP:journals/jmlr/ChangHCRL10}. For sparse features, Factorization Machine~(FM) and its variants project each feature into a low-dimensional vector and model cross features by inner product \cite{DBLP:conf/icdm/Rendle10}. 
SFM introduces Laplace distribution to model the parameters and better fit the sparse data with a higher ratio of zero elements~\cite{pan2016sparse}. FFM enables each feature to have multiple latent vectors to interact with features from different fields \cite{juan2016field}.
As FM and its variants can only model order-2nd cross features. An efficient algorithm Higher-Order FM (HOFM) for training arbitrary-order cross features was proposed by introducing the ANOVA kernel~\cite{blondel2016higher}. As reported in \cite{DBLP:conf/ijcai/XiaoY0ZWC17}, HOFM achieves marginal improvement over FM whereas using many more parameters and only its low-order (usually less than 5) form can be practically used.

\textbf {Implicit deep learning models.} As deep learning has shown promising representation capabilities, several models use deep learning to improve FM. 
Attention FM (AFM) enhances the importance of different order-2nd cross features via attention networks \cite{DBLP:conf/ijcai/XiaoY0ZWC17}. 
Neural FM (NFM) stacks deep neural networks on top of the output of the order-2nd cross features to model higher-order cross features~\cite{he2017neural}.
FNN uses FM to pre-train low-order features and then feeds feature embeddings into an MLP \cite{zhang2016deep}.
In contrast, DSL uses MLP to pre-train high-order non-linear features and then feeds them with basis features into an FM layer~\cite{huang2017ad}.
Moreover, CCPM uses convolutional layers to explore local-global dependencies of cross features \cite{liu2015convolutional}.
IPNN (also known as PNN) feeds the interaction results of the FM layer and feature embeddings into an MLP \cite{qu2016product}. PIN introduces a micro-network for each pair of fields to model pairwise cross features \cite{qu2018product}. 
FGCNN combines a CNN and MLP to generate new features for feature augmentation~\cite{liu2019feature}.
However, all these approaches learn the high-order cross features in an implicit manner, therefore lack good model explainability.

\textbf {Wide\&Deep based models.}  Jianxun Lian et al. argue that implicit deep learning models focus more on high-order cross features but capture little low-order cross features \cite{lian2018xdeepfm}. 
To overcome this problem, there has been proposing a hybrid network structure, namely Wide\&Deep, which combines a shallow component and a deep component with the purpose of learning both memorization and generalization \cite{cheng2016wide}.
Wide\&Deep framework revolutionizes the development of CTR prediction, and attracts industry partners a lot from the beginning.
As for the first Wide\&Deep model proposed by Google \cite{cheng2016wide}, it combines a linear model (wide part) and DNN, while the input of the wide part still relies on feature engineering. 
Later on, DeepFM uses an FM layer to replace the wide component. Deep\&Cross \cite{wang2017deep} and xDeepFM~\cite{lian2018xdeepfm} take outer product of features at the bit- and vector-wise level respectively. However, xDeepFM uses so many parameters that great challenges are posed to identify important cross features in the huge combination space.

 \textbf {AutoML based models.} There exist some pre-trained approaches using AutoML techniques to deal with cross features. AutoCross is proposed to search over subsets of candidate features to identify effective interactions \cite{luo2019autocross}. This requires training the whole model to evaluate the selected feature interactions, but the candidate sets are incredibly many. AutoGroup treats the selection process of high-order feature interactions as a structural optimization problem, and solves it with Neural Architecture Search \cite{DBLP:conf/sigir/LiuXGTZHL20}. It achieves state-of-the-art performance on various datasets, but is too complex to be applied in industrial applications.

\section{Extreme Cross Network (XCrossNet)}

\begin{figure}[t]
	\centering
	\includegraphics[width=0.78\textwidth,height=5.7cm]{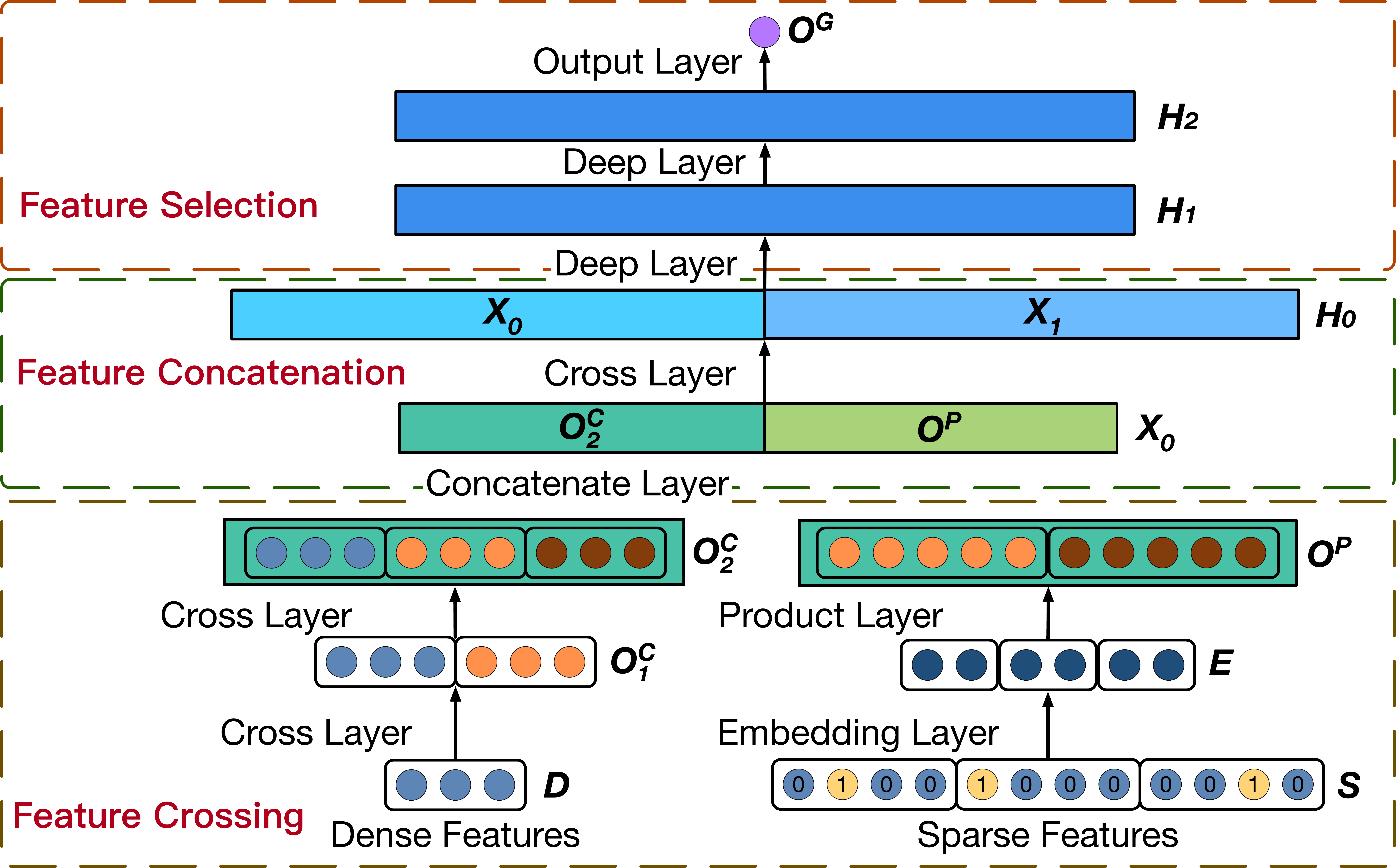}
	\caption{The structure of XCrossNet.} \label{fig3}
\end{figure}

In this section, we will introduce the problem statement and describe the details of Extreme Cross Network (XCrossNet) in the following three steps: Feature Crossing, Feature Concatenation, and Feature Selection. The complete XCrossNet model is depicted in Figure \ref{fig3}.

\subsection{Problem Statement}

In web-scale commercial recommender systems, the inputs of users' characteristics are in two kinds of structures. The first kind of structure is described by numerical or dense parameters, denoted as $\bm{D}$. The second kind of structure is described by categorical or sparse parameters, denoted as $\bm{S}$.
Suppose that the dataset for training consists of $n$ instances $([\bm{D}; \bm{S}], y)$, where $\bm{D} = [D_1, D_2, \cdots, D_M]$ indicates dense features including $M$ numerical fields, and $\bm{S} = [S_1, S_2, \cdots, S_N]$ indicates sparse features including $N$ categorical fields, and $y \in \{0, 1\}$ indicates the user's click behaviors ($y=1$ means the user clicked the item, and $y=0$ otherwise). The task of CTR prediction is to build a prediction model $\hat{y}=pCTR\_Model([\bm{D}; \bm{S}])$ to estimate the ratio of clicks to impressions of a given feature context.

\subsection{Feature Crossing}

A cross feature is defined as a synthetic feature formed by multiplying (crossing) two features. Crossing combinations of features can provide predictive abilities beyond what those features can provide individually.
Based on the definition, cross features can be generalized to high-order cases. If we consider individual features as order-$1$st features, an order-$k$th cross feature is formed by multiplying $k$ individual features. 

\subsubsection{Cross layers on dense features.}
\begin{figure}[t]
	\centering
	\includegraphics[width=0.8\textwidth]{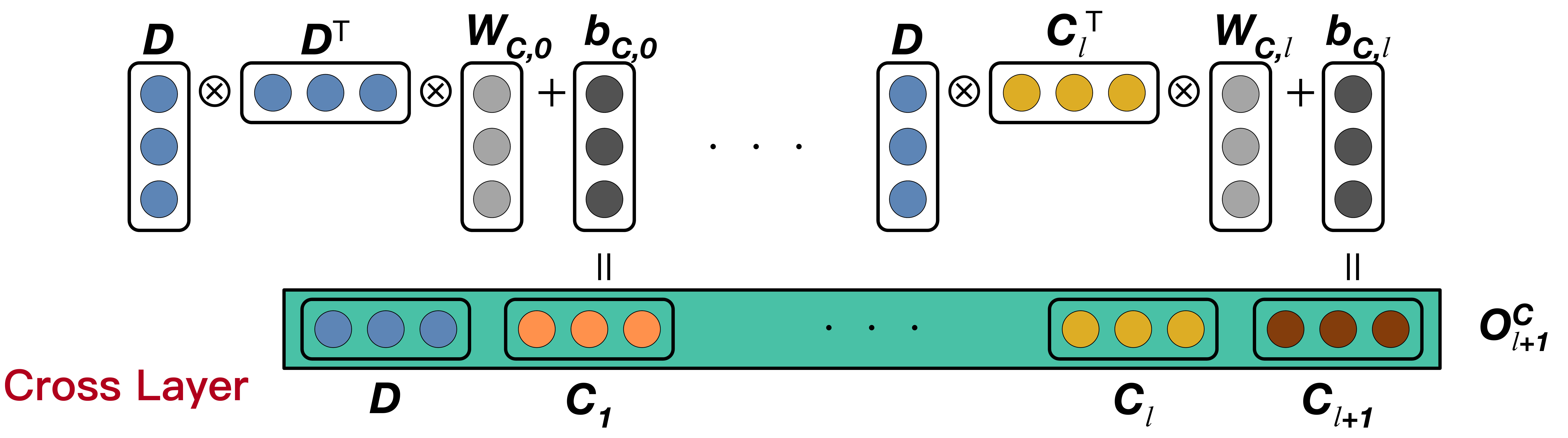}
	\caption{The structure of cross layers.} \label{fig1}
\end{figure}

First we introduce a novel cross layer for crossing dense features (see in Fig. \ref{fig1}). Cross layers have the following formula: 
\begin{small}
	\begin{equation}
	\begin{aligned}
	\bm{C_1} &= \bm{D} \cdot \bm{D^\mathsf{T}} \cdot \bm{W_{C,0}} + \bm{b_{C,0}}, \quad
	\bm{O^{C}_{1}} = [\bm{D}; \bm{C_1}], \\
	\bm{C_{l+1}} &= \bm{D} \cdot \bm{C_{l}^\mathsf{T}} \cdot \bm{W_{C,l}} + \bm{b_{C,l}}, \quad
	\bm{O^{C}_{l+1}} = [\bm{O^{C}_{l}}; \bm{C_{l+1}}],
	\end{aligned}
	\end{equation}
\end{small}where $\bm{D} \in \mathbb{R}^M$ indicates the input dense features, and $\bm{C_l} \in \mathbb{R}^M$ is a column vector denoting the order-$(l+1)$th cross features. Later we prove how $\bm{C_l}$ expresses multivariate polynomials of degree $(l+1)$ after weighted mapping. $\bm{W_{C,l}}, \bm{b_{C,l}} \in \mathbb{R}^M$ are the weight and bias parameters respectively, and $\bm{O^{C}_{l}}, \bm{O^{C}_{l+1}}$ denote the outputs from the $l$-th and the $(l+1)$-th cross layers.

We denote $\bm{\alpha} = [\alpha_{1}, \cdots, \alpha_{M}]$. If our proposed cross layer expresses any cross features of order-$(l+1)$th, it could approximate to any multivariate polynomials of degree $(l+1)$, denoted as $P_{l+1}(\bm{D})$: 
\begin{small}
	\begin{equation}
	P_{l+1}(\bm{D}) = \bigg \{ \sum_{\bm{\alpha}} W_{\bm{\alpha}} D_1^{\alpha_1} D_2^{\alpha_2} \cdots D_M^{\alpha_M} \, \bigg | \,|\bm{\alpha}|=l+1 \bigg \}, 
	\end{equation}\end{small}where $|\bm{\alpha}|=\sum_{i=1}^{M}\alpha_{i}$. For simplicity, here we use $\bm{W^{i}} = [W^{i}_1, W^{i}_2, \cdots, W^{i}_M]$ to denote the original subscript of $\bm{W_{C,i}}$. We study the coefficient $\hat{W_{\bm{\alpha}}}$ given by $\bm{C_l^{\mathsf{T}}} \cdot \bm{W^{l}}$ from cross layers, since it constitutes the output $\bm{O_{l+1}^{C}}$ from the $(l+1)$-th cross layer. Besides, the following derivations do not include bias terms. Then: 
\begin{small}
	\begin{equation}
	\begin{array}{l}
	\bm{C_l^{\mathsf{T}}} \cdot \bm{W^{l}} = \Big(\bm{C_{l-1}^{\mathsf{T}}} \cdot \bm{W^{l-1}} \Big) \cdot \Big( \bm{D^\mathsf{T}} \cdot \bm{W^{l}} \Big) = \prod_{i=0}^{l} \bm{D^{\mathsf{T}}} \cdot \bm{W^{i}} \\
	= \prod_{i=0}^{l} [D_1, D_2, \cdots, D_M]^{\mathsf{T}} \cdot [W^{i}_1, W^{i}_2, \cdots, W^{i}_M].
	\end{array}
	\end{equation}
\end{small}Afterwards, let $\bm{I}$ denotes the multi-index vectors of orders $[0, 1, \cdots, l]$, and $I_j$ denotes the order of field $j$. Clearly $\bm{C_l^{\mathsf{T}}} \cdot \bm{W^{l}}$ from cross layers approaches the coefficient $\hat{W_{\bm{\alpha}}}$~as: 
\begin{small}
	\begin{equation}
	\hat{W_{\bm{\alpha}}} = \sum_{k=1}^{M} \sum_{|\bm{I}|=\alpha_k} \prod_{j=1}^{M} W_{j}^{I_j}. 
	\end{equation}
\end{small}With $\bm{C_l^{\mathsf{T}}} \cdot \bm{W^{l}}$ approximate to multivariate polynomials of degree $(l+1)$, the output $\bm{O^{C}_{l+1}}$ from the $(l+1)$-th cross layer that includes all cross features to order-$(l+1)$th could approximate polynomials in the following class: 
\begin{small}
	\begin{equation}
	P_{l+1}(\bm{D}) = \bigg \{ \sum_{\bm{\alpha}} W_{\bm{\alpha}} D_1^{\alpha_1} D_2^{\alpha_2} \cdots D_M^{\alpha_M} \, \bigg | \,0 \le |\bm{\alpha} | \le l+1 \bigg \}. 
	\end{equation}
\end{small}\subsubsection{Embedding and product layers on sparse features.}
\begin{figure}[t]
	\centering
	\includegraphics[width=0.8\textwidth]{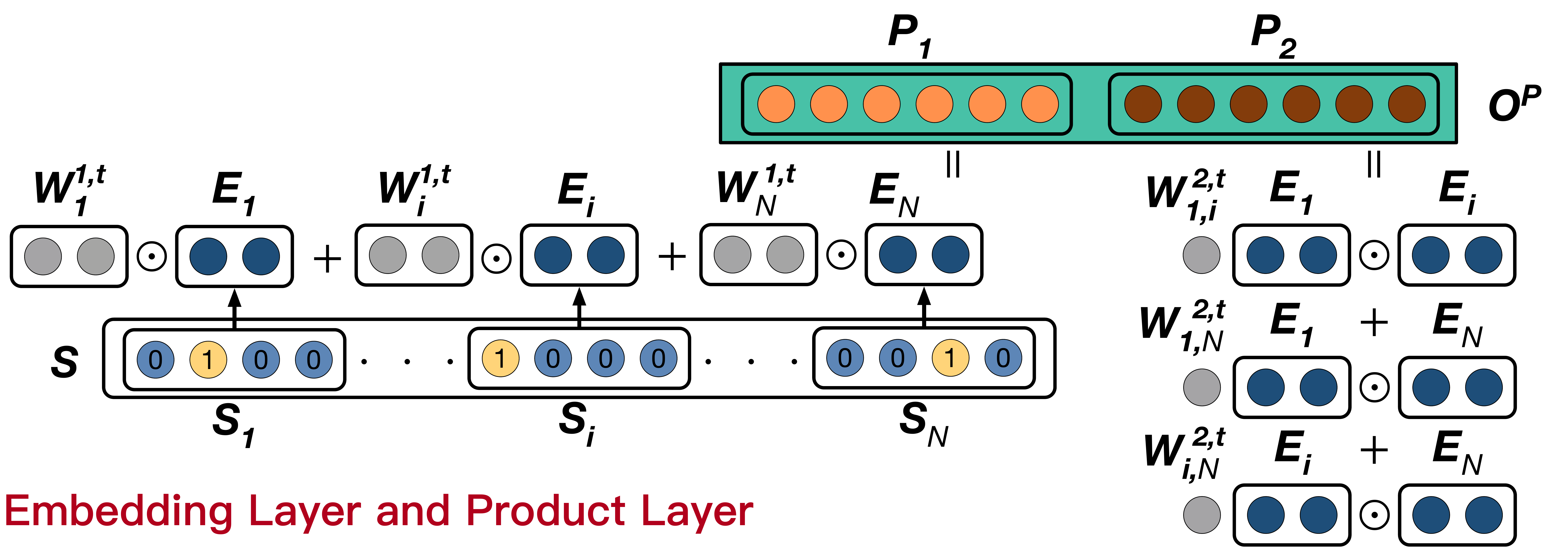}
	\caption{The structure of embedding layer and product layer.} \label{fig2}
\end{figure}

Here we introduce the embedding layer and product layer for crossing sparse features (see in Fig.~\ref{fig2}).
As sparse features $\bm{S}$ are represented as vectors of one-hot encoding of high-dimensional spaces, we employ an embedding layer to transform these one-hot encoding vectors into dense vectors $\bm{E}$ as:

\begin{small}
	\begin{equation}
	\begin{aligned}
	\bm{E} &= [\bm{E_1}, \cdots, \bm{E_i}, \cdots, \bm{E_N}], \\
	\bm{E_i} &= {\rm{embed}} (\bm{S_i}), \, \big( \bm{E_i} \in \mathbb{R}^K, i=1, \cdots, N \big)
	\end{aligned}
	\end{equation}
\end{small}where $\bm{S_i}$ indicates the input sparse feature of field $i$, $K$ denotes the embedding size, and $\bm{E_i}$ denotes the feature embedding of field~$i$.

Afterwards, we can propose a product layer for cross sparse features. First, we donate order-2nd cross sparse features as $\bm{P_2}$, and order-1st sparse features as $\bm{P_1}$, thus the output of product layer is $\bm{O^P} = [\bm{P_1}; \bm{P_2}]$.

The cross feature of two sparse features of field $i$ and field $j$ equals the inner product of two embedding vectors as $\langle \bm{E_i}, \bm{E_j} \rangle$. Intuitively, we expect cross features to be vectors, so we concatenate the weighted sums of inner products to formulate order-2nd cross features as: 
\begin{small}
	\begin{equation}
	\bm{P_2} = [P_2^1, \cdots, P_2^t, \cdots, P_2^T], 
	\end{equation}
\end{small}where $T$ is the size of product layer, and $\bm{P_2}$ is a $T$ dimensional vector, of each dimension $P_2^t$ denotes a weighted sum of inner products of two sparse features. Thus, we have $P_2^t = \sum_{i=1}^{N} \sum_{j=1}^{N} W^{2,t}_{i,j} \langle \bm{E_i}, \bm{E_j} \rangle $. We assume that the weighted parameter $W^{2,t}_{i,j} = \Theta_{i}^{t} \cdot \Theta_{j}^{t}$ for reduction, so $P_2^t$ can be given as: 
\begin{small}
	\begin{equation}
	P_2^t= \sum_{i=1}^{N} \sum_{j=1}^{N} \Theta_{i}^{t} \cdot \Theta_{j}^{t} \langle \bm{E_i}, \bm{E_j} \rangle
	= \bigg\langle \sum_{i=1}^{N} \Theta_{i}^{t} \cdot \bm{E_i}, \sum_{j=1}^{N} \Theta_{j}^{t} \cdot \bm{E_j} \bigg\rangle.
	\label{eq1}
	\end{equation}
\end{small}The feature vector of order-1st features has a similar formula as follows: 
\begin{small}
	\begin{equation}
	\bm{P_1} = [P_1^1, \cdots, P_1^t, \cdots, P_1^T],
	\end{equation}
\end{small}where $\bm{P_1}$ is a $T$ dimensional vector, of each dimension $P_1^t$ denotes a weighted sum of sparse features. The weighted feature can be expressed as inner product $\langle \bm{W^{1,t}_{i}}, \bm{E_i} \rangle$. Thus, we have $P_1^t = \sum_{i=1}^{N} \langle \bm{W^{1,t}_{i}}, \bm{E_i} \rangle$.

\subsection{Feature Concatenation}
In the Feature Concatenation stage, in order to learn feature interactions of different structures, cross dense features $\bm{O^C}$ and cross sparse features $\bm{O^P}$ are concatenated as a vector through a concatenate layer, then the concatenated feature vector is fed into a cross layer, which can be expressed as:
\begin{small}
	\begin{equation}
	\begin{aligned}
	\bm{X_0} &= [\bm{O^C}; \bm{O^P}], \\
	\bm{X_1} &= \bm{X_0} \cdot \bm{X_0^\mathsf{T}} \cdot \bm{W_{X,0}} + \bm{b_{X,0}},  \quad
	\bm{H_0} = [\bm{X_0}; \bm{X_1}],
	\end{aligned}
	\label{eq2}
	\end{equation}
\end{small}where $\bm{X_0}$ denotes the concatenated feature of cross dense features and cross sparse features, $\bm{X_1}$ denotes the cross features between two kinds of feature structures, $\bm{H_0}$ denotes the output from this cross layer, and $\bm{W_{X,0}}, \bm{b_{X,0}}$ are the weight and bias parameters of this cross layer.

\subsection{Feature Selection}
In the Feature Selection stage, we employ an MLP to capture non-linear interactions and the relative importance of cross features. The deep layers and the output layer respectively have the following formula: 
\begin{small}
	\begin{equation}
	\begin{aligned}
	\bm{H_i} &= {\rm{ReLU}} ( \bm{W_{H,i-1} } \cdot \bm{H_{i-1}} + \bm{ b_{H,i-1} } ), \\
	O^G &= {\rm{Sigmoid}} (\bm{W_{H,i}} \cdot \bm{H_i} + \bm{b_{H,i}} ),
	\end{aligned}
	\end{equation}
\end{small}where $\bm{H_i}, \bm{H_{i-1}}$ are hidden layers, ${\rm{ReLU}} (\cdot)$ and ${\rm{Sigmoid}}(\cdot)$ are activation functions, $\bm{W_{H,i}}, \bm{W_{H,i-1} }$ are weights, and $\bm{b_{H,i}}, \bm{ b_{H,i-1} }$ are biases, and $O^G$ is the output result.

For CTR prediction, the loss function is the Logloss as follows: 
\begin{equation}
\bm{\mathcal{L}} = - \frac{1}{n} \sum^n_{i=1} \, y_i \log(O^G) + (1-y_i) \log(1-O^G),
\end{equation}where $n$ is the total number of training instances. The optimization process is to minimize the following objective function:
\begin{equation}
\bm{\mathcal{J}} = \bm{\mathcal{L}} + \lambda ||\bm{\Theta}||,
\end{equation}where $\lambda$ denotes the regularization term, and $\bm{\Theta}$ denotes the set of learning parameters, including cross layers, embedding layer, product layer, deep layers and output layer.

\section{Experiments}

In this section, extensive experiments are conducted to answer the following research questions\footnote{We release the source code at https://github.com/bigdata-ustc/XCrossNet/.}:
\begin{description}
	\item[RQ1:] How does XCrossNet perform compared with the state-of-the-art CTR prediction models?
	\item[RQ2:] How does the feature dimensionality imbalance impact CTR prediction?
	\item[RQ3:] How do hyper-parameter settings impact the performance of XCrossNet?
\end{description}

	\newfloat{figtab}{htb}{fgtb}
	\makeatletter
	\newcommand\figcaption{\def\@captype{figure}\caption}
	\newcommand\tabcaption{\def\@captype{table}\caption}
	\makeatother

		\begin{figtab}
			\begin{minipage}[h]{0.45\linewidth}
				\centering
				\begin{tabular}{c|cc}
					\hline
					\hline
					Model & AUC(\%)  & Logloss    \\
					\hline
					LR                      & 78.00                    & 0.5631                         \\
					GBDT                      & 78.62                  & 0.5560                  \\
					FM                      & 79.09                  & 0.5500                    \\
					AFM                      & 79.13                    & 0.5517                                 \\
					FFM                      & 79.80                    & 0.5438                          \\
					%	\hline
					CCPM                      & 79.55                   & 0.5469                        \\
					%	\hline
					Wide\&Deep                      & 79.77                    & 0.5446              \\
					%	\hline
					Cross                      & 78.70                    & 0.5550                       \\
					Deep\&Cross                      & 79.76                   & 0.5445       \\
					FNN                      & 79.87                    & 0.5428                       \\
					%	\hline
					DeepFM                    & 79.91                    & 0.5423                     \\
					%	\hline
					IPNN                      & 80.13                    & 0.5399                       \\
					%	\hline
					PIN                      & 80.18                    & 0.5394                          \\
					%	\hline
					CIN                     & 78.81                   & 0.5538                            \\
					%	\hline
					xDeepFM                      & 80.06                    & 0.5408                   \\
					%	\hline
					FGCNN                      & 80.22                    & 0.5389                  \\
					%	\hline
					AutoGroup                      & 80.28                    & 0.5384              \\
					\hline
				XCrossNet                     & \textbf{80.68}      & \textbf{0.5339}       \\
					\hline
					\hline
				\end{tabular}\label{table1}

		\tabcaption{Performance comparison of different CTR prediction models.}
			\end{minipage}\quad
			\begin{minipage}[h]{0.5\linewidth}
				\begin{center}
				\includegraphics[width = 0.9\linewidth]{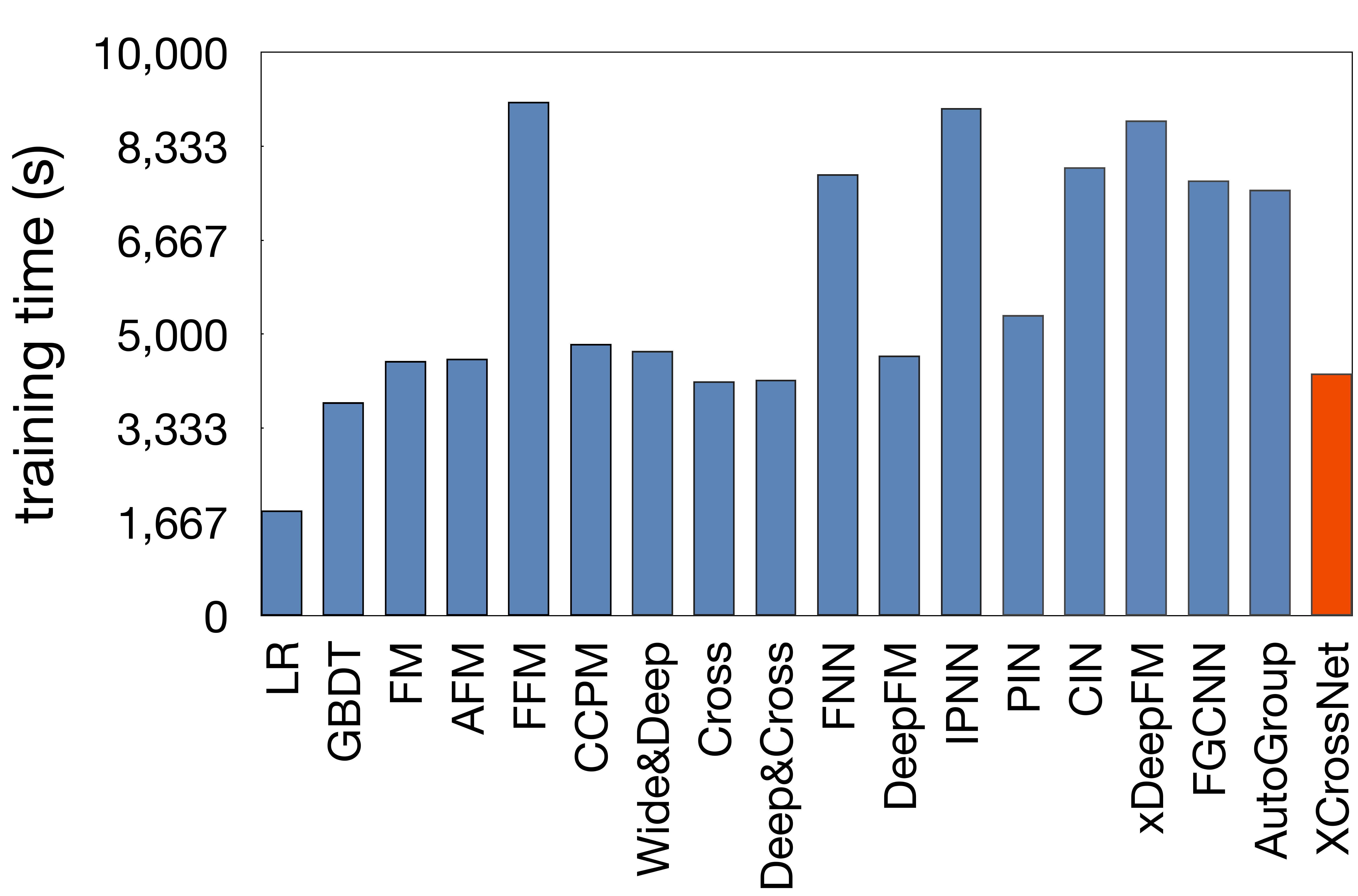}
				\figcaption{Training time comparison of different CTR prediction models.}\label{fig4}
				\end{center}
			\begin{center}
				\includegraphics[width = 0.9\linewidth]{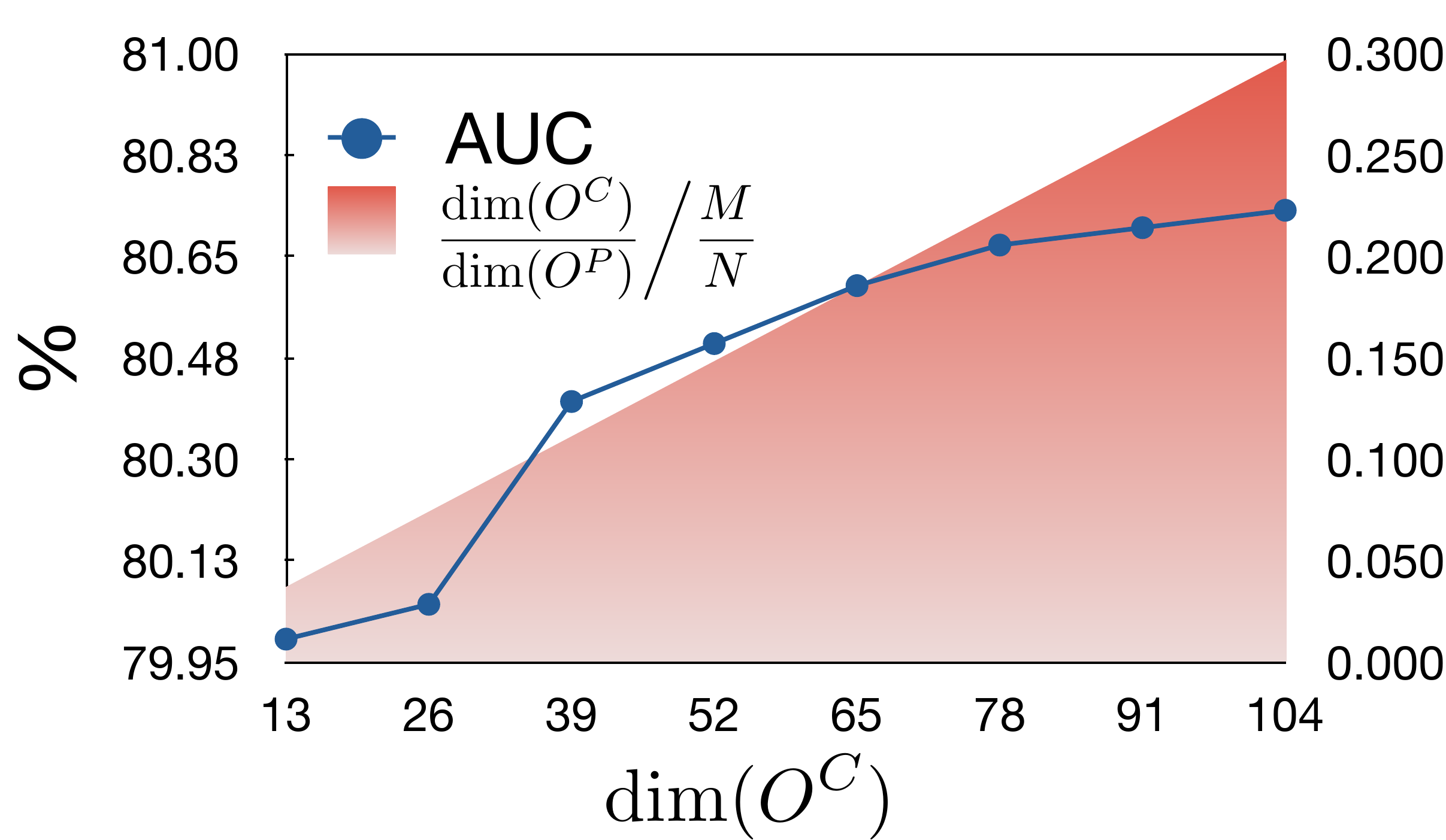}
				\figcaption{Impact of feature dimensionality imbalance.}\label{fig5}
			\end{center}
			\end{minipage}
		\end{figtab}
	
\subsection{Experimental Setup}
	\subsubsection{Dataset.} Experiments are conducted on Criteo Kaggle dataset, which is from a world-wide famous Demand-Side Platforms. Criteo Kaggle dataset contains one month of $45,840,617$ ad click instances. It has $13$ integer feature fields and $26$ categorical feature fields.  We select 7 consecutive days of samples as the training set while the next one day for evaluation. 

	\subsubsection{Baselines.} As aforementioned, we use following highly related state-of-the-art models as baselines: \textbf{LR} \cite{DBLP:conf/kdd/LeeODL12}, \textbf{GBDT} \cite{friedman2001greedy}, \textbf{FM} \cite{DBLP:conf/icdm/Rendle10}, \textbf{AFM} \cite{DBLP:conf/ijcai/XiaoY0ZWC17}, \textbf{FFM} \cite{juan2016field}, \textbf{CCPM}~\cite{liu2015convolutional}, \textbf{Wide\&Deep} \cite{cheng2016wide}, \textbf{Deep\&Cross} \cite{wang2017deep} and its shallow part \textbf{Cross} network, \textbf{FNN} \cite{zhang2016deep}, \textbf{DeepFM} \cite{DBLP:conf/ijcai/GuoTYLH17}, \textbf{IPNN} \cite{qu2016product}, \textbf{PIN} \cite{qu2018product}, \textbf{xDeepFM} \cite{lian2018xdeepfm} and its shallow part \textbf{CIN}, \textbf{FGCNN} \cite{liu2019feature}, and \textbf{AutoGroup} \cite{DBLP:conf/sigir/LiuXGTZHL20}.

\subsubsection{Hyper-parameter settings.} 
	 For model optimization, we use Adam with a mini-batch size of 4096, and the learning rate is set as $0.001$. 
	 We use the L2 regularization with $\lambda = 0.0001$ for all neural network models. 
	  For Wide\&Deep, Deep\&Cross, FNN, DeepFM, IPNN, PIN, xDeepFM, and XCrossNet, the numbers of neurons per deep layer are $400$, and the depths of deep layers are set as~$2$.
	 For our XCrossNet, the number of cross layers on dense features is set as $l$=4. 
	  In the main experiments, we set the embedding size for all models be a fixed value of $20$.

\subsection{Overall Performance (RQ1)}
Table~\ref{table1} summarizes the performance of all compared methods on Criteo Kaggle datasets, while the training time on Tesla K80 GPUs is shown in Fig. \ref{fig4} for comparison of efficiency. From the experimental results, we have the following key observations:
Firstly, most neural network models outperform linear models (i.e., LR), tree-based models (i.e., GBDT), and FM variants (i.e., FM, FFM, AFM), which indicates MLP can learn non-linear feature interactions and endow better expressive ability. Meanwhile, comparing IPNN, PIN with FNN, Wide\&Deep based models, we find that explicitly modeling low-order feature interactions can simplify the training of MLP and boost the performance.
Secondly, XCrossNet achieves the best performance. Statistically, XCrossNet significantly outperforms the best baseline in terms of AUC and Logloss on $p$-value $<0.05$ level, which indicates feature structure-oriented learning can provide better predictive abilities.
Thirdly, from the training time comparison, we can observe XCrossNet is very efficient, especially compared to field-aware models, mainly because these models further allow each feature to learn several vectors where each vector is associated with a field, which leads to huge parameter consumption and time consumption.

\begin{figure}[!t]
	\centering
	
	\subfigure[]{
		\begin{minipage}[t]{3.2cm}
			\centering
			\includegraphics[width=3.3cm,height=2.5cm]{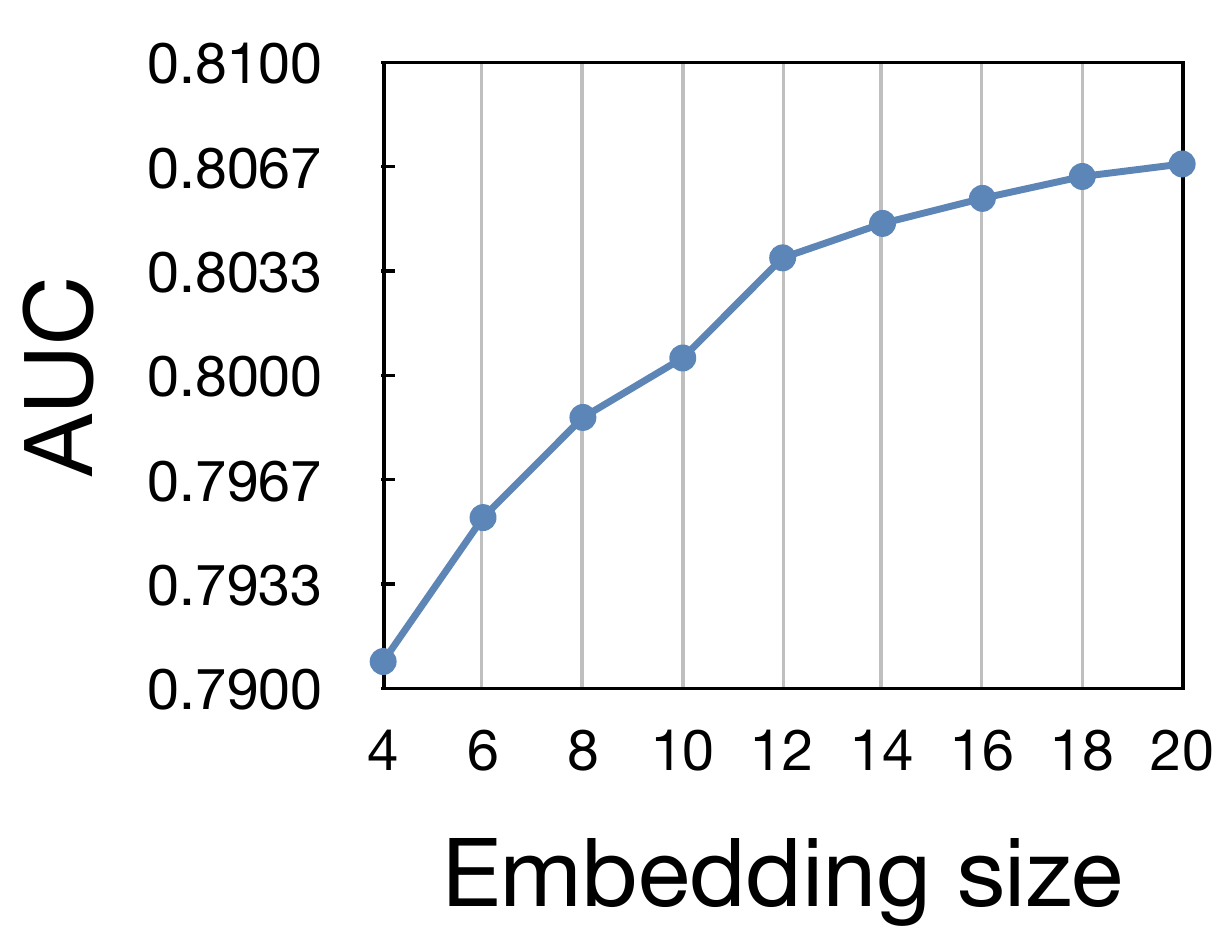}
		\end{minipage}%
	}%
	\subfigure[]{
		\begin{minipage}[t]{2.9cm}
			\centering
			\includegraphics[width=3cm,height=2.5cm]{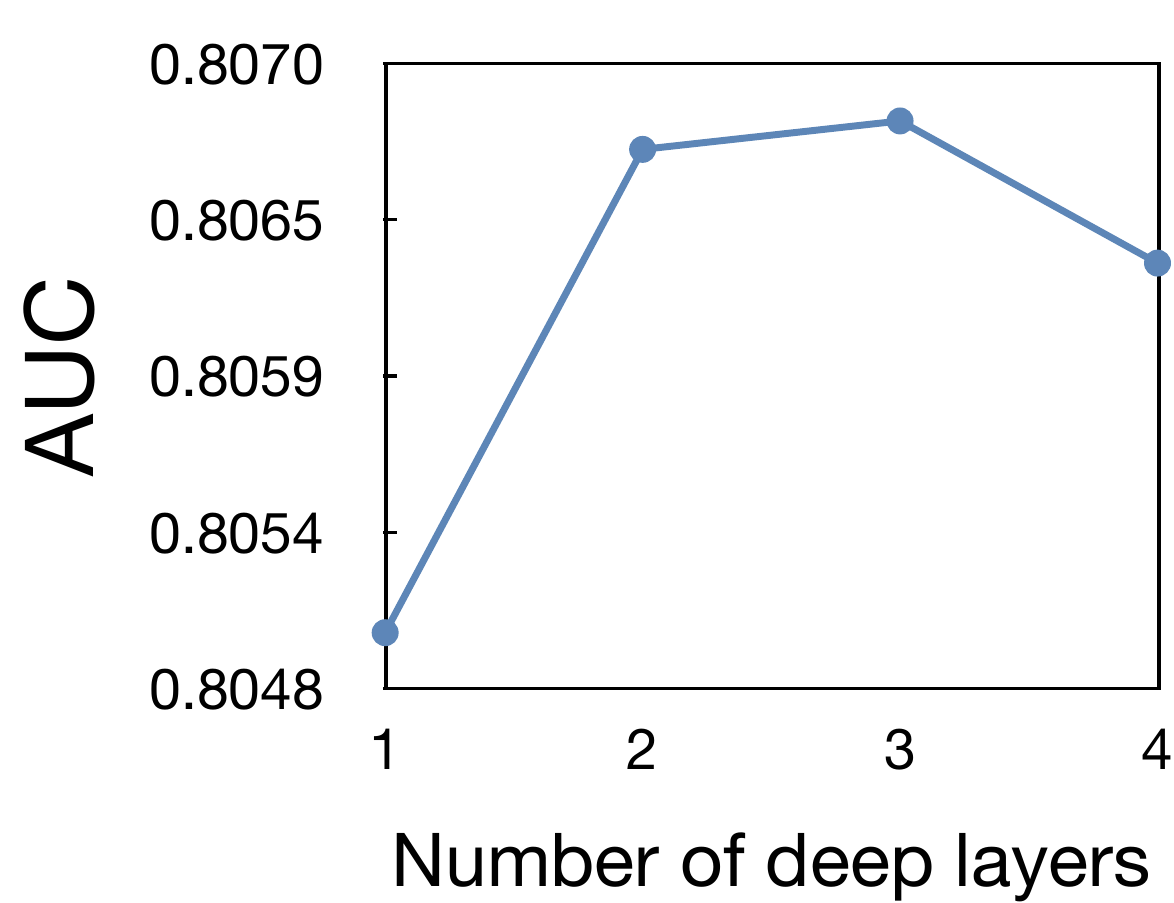}
		\end{minipage}%
	}%
	\subfigure[]{
		\begin{minipage}[t]{2.8cm}
			\centering
			\includegraphics[width=2.9cm,height=2.5cm]{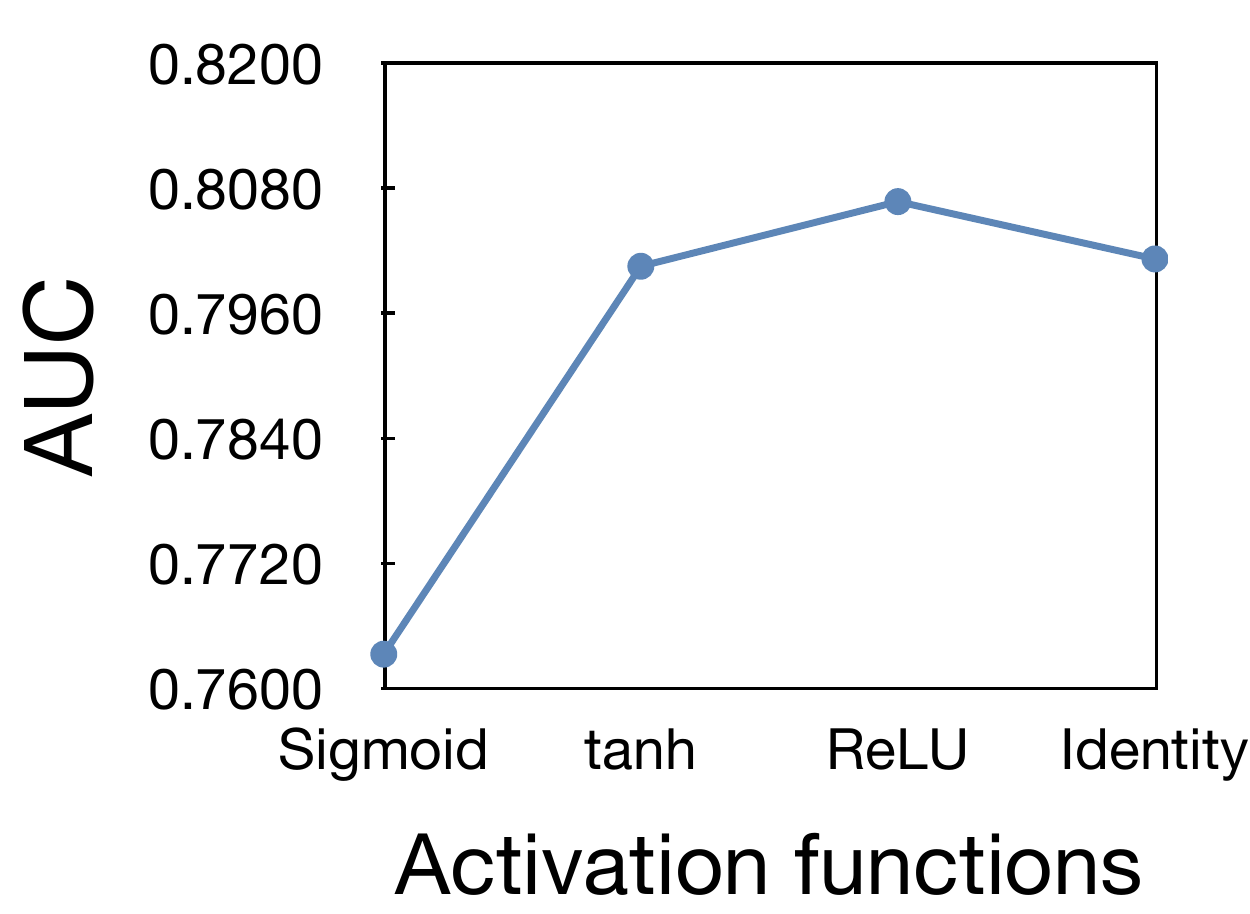}
		\end{minipage}%
	}%
	\subfigure[]{
		\begin{minipage}[t]{2.8cm}
			\centering
			\includegraphics[width=2.9cm,height=2.5cm]{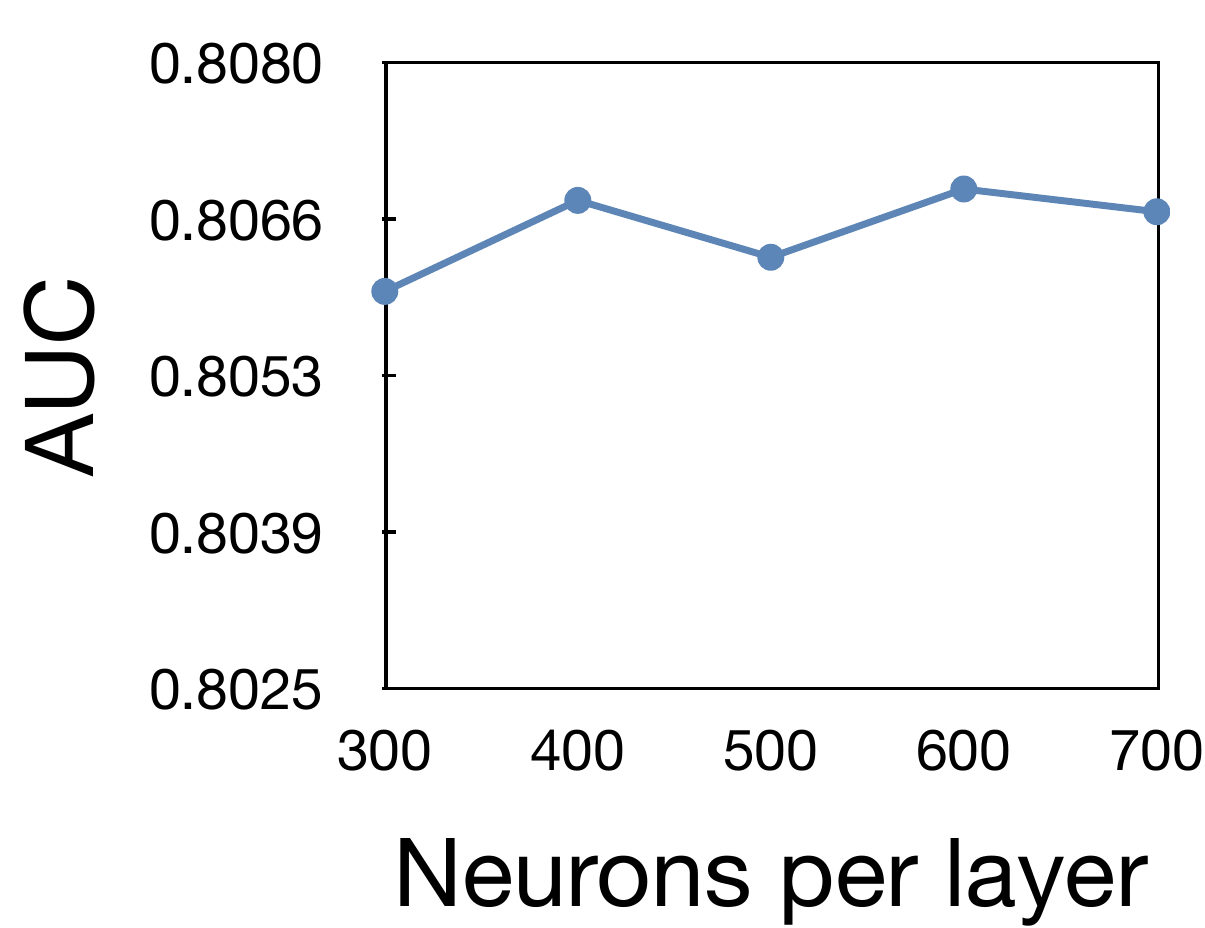}
		\end{minipage}%
	}%
	
	\centering
	\caption{Impact of network hyper-parameters on AUC performance.}
	\label{fig6}
\end{figure}
\begin{figure}[t]
	\centering
	
	\subfigure[]{
		\begin{minipage}[t]{3.2cm}
			\centering
			\includegraphics[width=3.3cm,height=2.5cm]{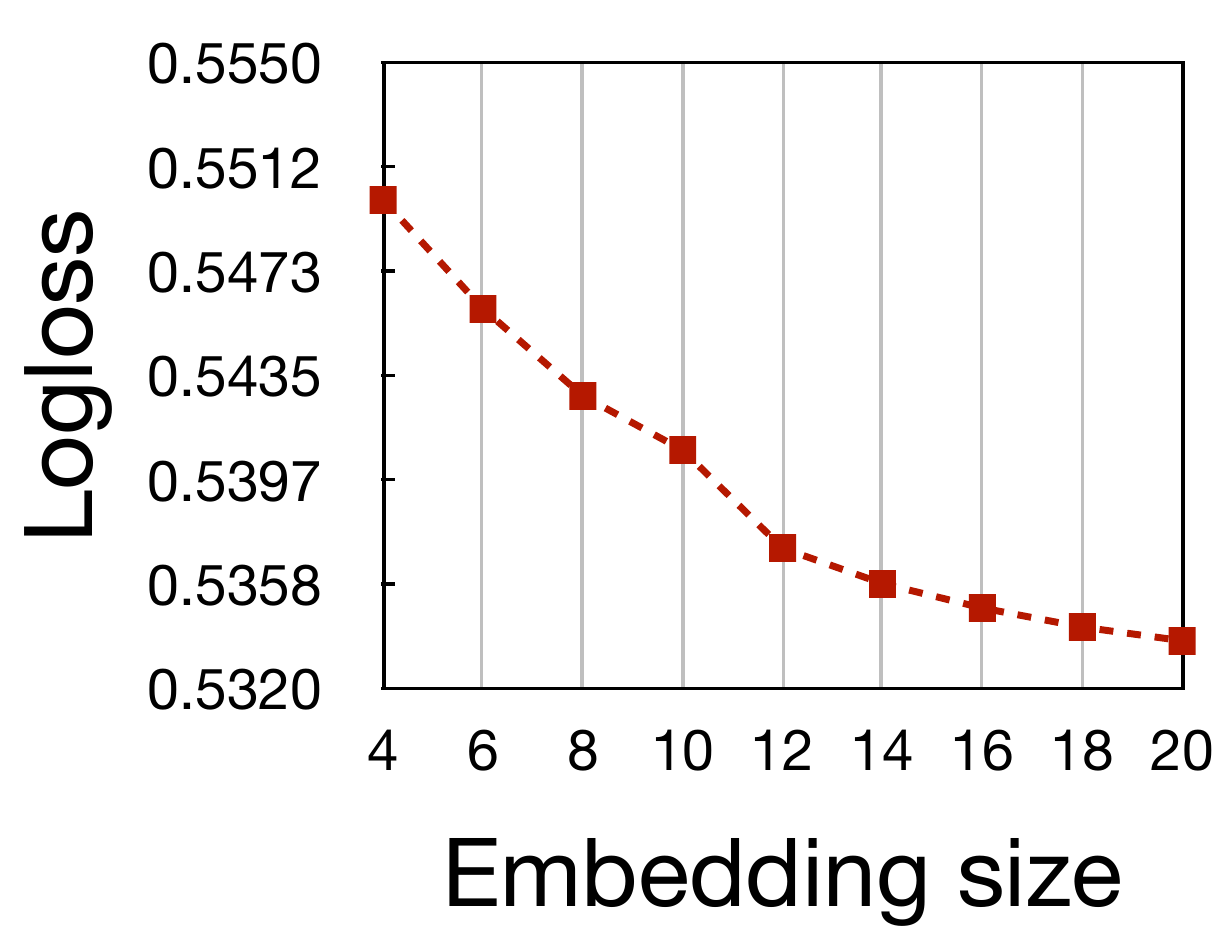}
		\end{minipage}%
	}%
	\subfigure[]{
		\begin{minipage}[t]{2.9cm}
			\centering
			\includegraphics[width=3cm,height=2.5cm]{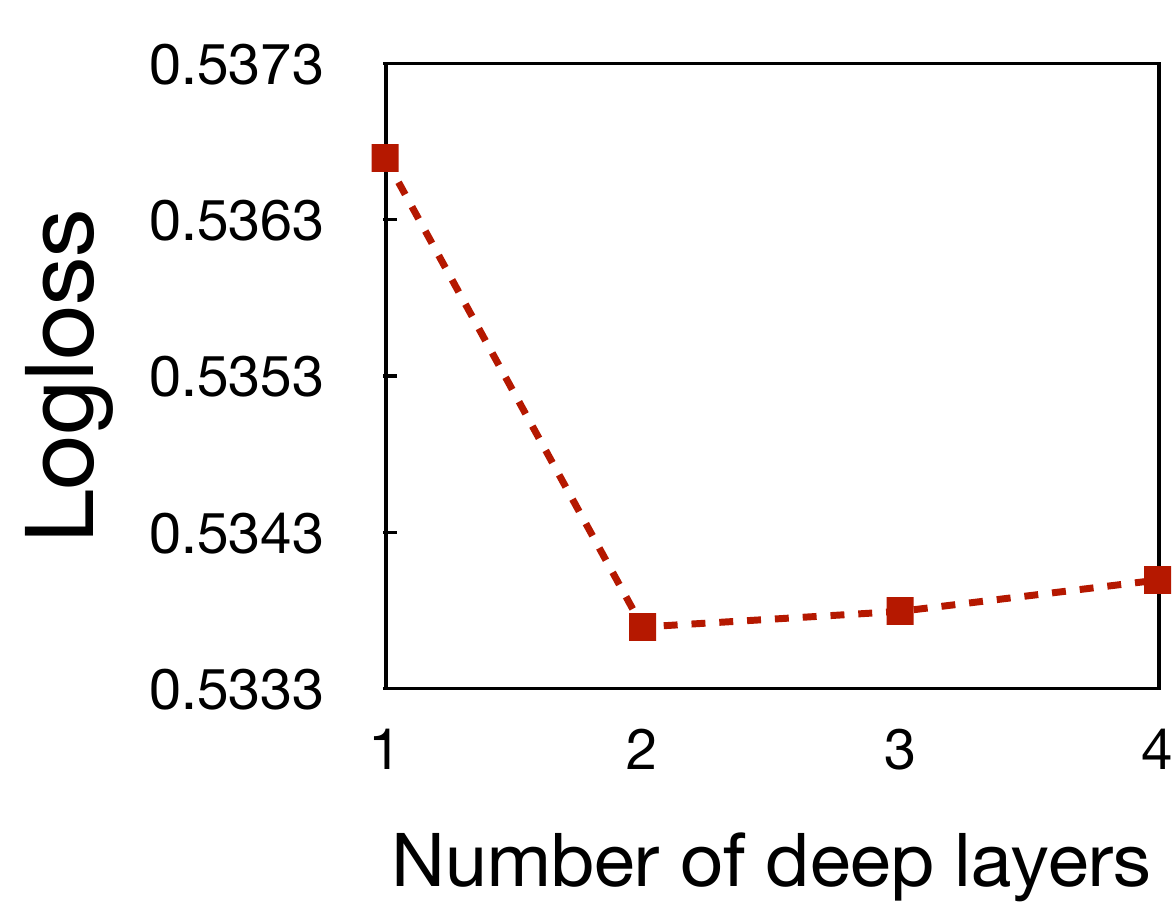}
		\end{minipage}%
	}%
	\subfigure[]{
		\begin{minipage}[t]{2.8cm}
			\centering
			\includegraphics[width=2.9cm,height=2.5cm]{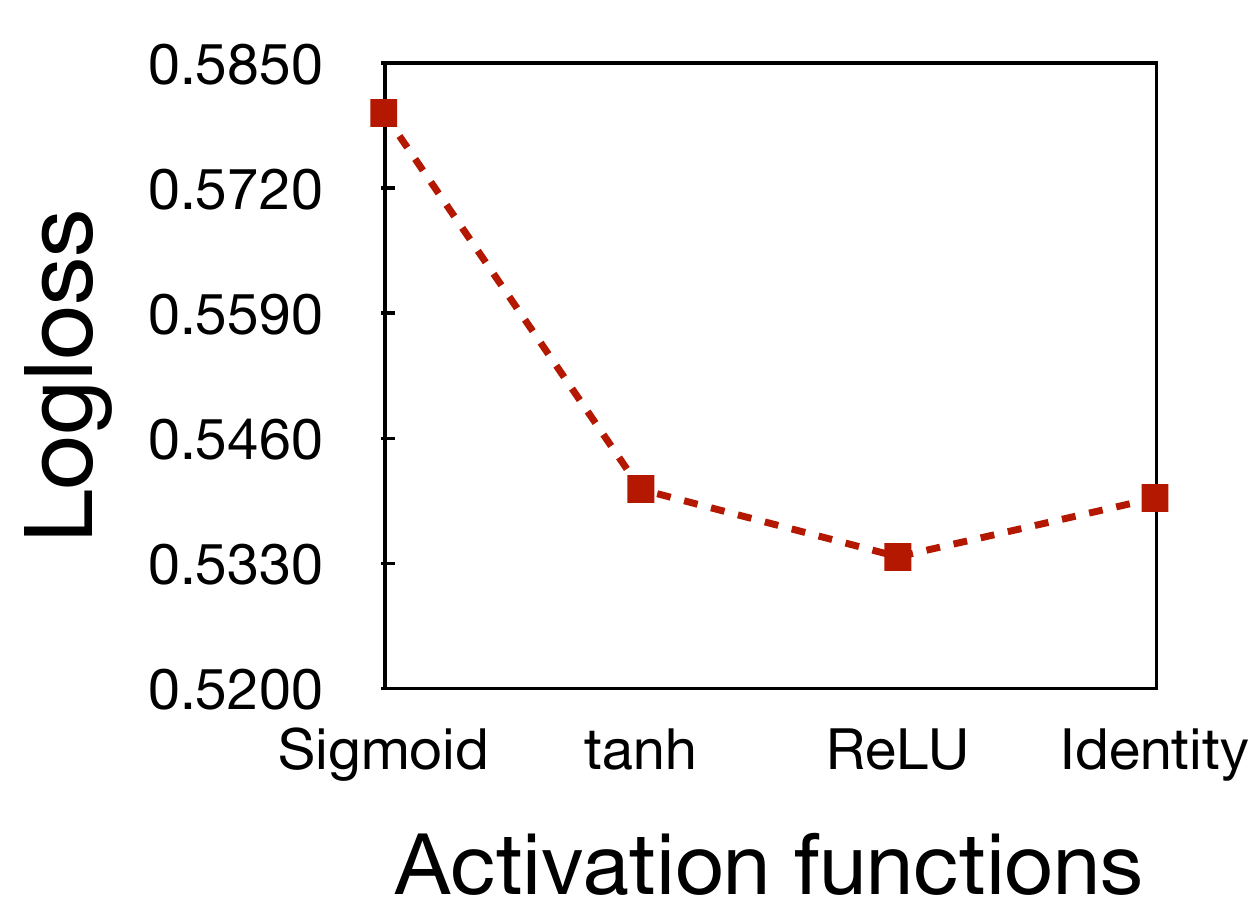}
		\end{minipage}%
	}%
	\subfigure[]{
		\begin{minipage}[t]{2.8cm}
			\centering
			\includegraphics[width=2.9cm,height=2.5cm]{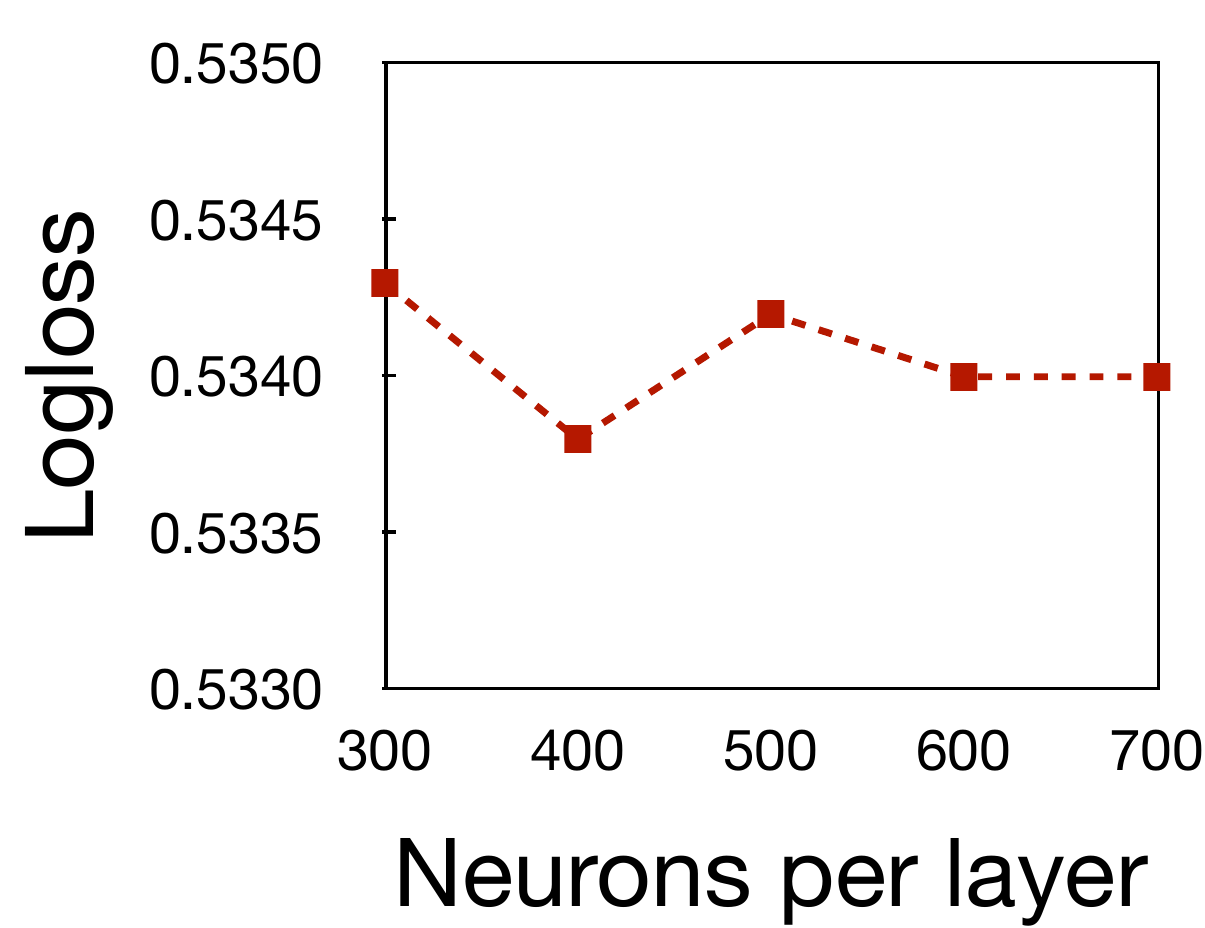}
		\end{minipage}%
	}%
	\centering
	\caption{Impact of network hyper-parameters on Logloss performance.}
	\label{fig7}
\end{figure}

\subsection{Feature Dimensionality Imbalance Study (RQ2)}
In XCrossNet, we denote $\left.{  \frac{\dim (O^C)} {\dim (O^P) } } \middle/ { \frac{M}{N} }\right.$ as the balance index of dimensions of dense and sparse features.
Noted that, the dimension of cross dense features $O^C$ equals $M\cdot l$, increasing with the depth of cross layers. As for Criteo Kaggle dataset, $M=13$ and $N=26$, we set the depths of cross layers from 1 to 8, while the corresponding dimensions of cross dense features are from 13 to 104. Experimental results are shown in Fig. \ref{fig5} in terms of AUC. We can observe that increasing the depth of cross layers benefits XCrossNet to achieve stable improvements on AUC performance, mainly because the higher dimensions of cross dense features are able to boost the balance index, which results in relatively balanced impacts of dense and sparse features on prediction.

\subsection{Hyper-parameter Study (RQ3)}
We study the impact of hyper-parameters of XCrossNet, including (1) embedding size; (2) number of deep layers; (3) activation function; (4) neurons per layer.
Figures 6a and 7a demonstrate the impact of embedding size. We can observe that model performance boosts steadily when the embedding size increase from 4 to 20. Even with very low embedding sizes, XCrossNet still has comparable performance to some popular Wide\&Deep based models with high embedding size. Specifically, XCrossNet achieves AUC$>0.800$ and Logloss$<0.541$ with embedding size set as $10$, which is even better than DeepFM with embedding size set as $20$. 
Figures 6b and 7b demonstrate the impact of the number of deep layers. The model performance boosts with the depth of MLP at the beginning. 
However, it starts to degrade when the depth of MLP is set to greater than 3.
As shown in Figures 6c and 7c, ReLU is indeed more appropriate for hidden neurons of deep layers compared with different activation functions.
As shown in Figures 6d and 7d, model performance barely boosts as the number of neurons per layer increasing from $300$ to $700$. 
We consider $400$ is a more suitable setting to avoid the model being overfitting.

\section{Conclusion}
Due to the fact that previous work rarely attempts to individually learn representations for different feature structures, this paper presented a novel feature structure-oriented learning model, namely Extreme Cross Network (XCrossNet), for improving CTR prediction in recommender systems. 
A XCrossNet model starts with a Feature Crossing stage, followed by a Feature Concatenation stage and a Feature Selection stage.
The main contribution of our approach is to represent dense and sparse feature interactions in an explicit and efficient way.
Empirical studies verified the effectiveness of our model on Criteo Kaggle dataset.

\subsubsection{Acknowledgements.} This research was partially supported by grants from the National Key Research and Development Program of China (No. 2018YFC0832101), and the National Natural Science Foundation of China (Grants No. 61922073 and U20A20229). Qi Liu acknowledges the support of the Youth Innovation Promotion Association of CAS (No. 2014299).

\bibliographystyle{splncs04}
\bibliography{ref4.bib}

\end{document}